\documentclass[12pt]{article} 
\usepackage{amssymb,overcite} 
 
\newcommand{\comment}[1]{} 
 
\begin{document}

\title{Relationship between quantum decoherence times and 
solvation dynamics in condensed 
phase chemical systems} 
 
\author{Oleg V. Prezhdo\thanks{
Permanent Address: Department of Chemistry, University 
of Washington, Seattle, WA 98195}
~and Peter J. Rossky\\ \\  
\em Department of Chemistry and Biochemistry\\ 
\em University of Texas at Austin\\ 
\em Austin, Texas 78712-1167} 
 
\date{Submitted to Phys. Rev. Lett.: \today} 
 
\maketitle 
 
 
\begin{abstract} 
\setlength{\baselineskip}{7mm}     
A relationship between the time scales of quantum coherence loss and 
short-time solvent response for a solute/bath system 
is derived for a Gaussian wave packet approximation 
for the bath. Decoherence and solvent 
response times are shown to be directly proportional to each other, with the 
proportionality coefficient given by the ratio of the thermal 
energy fluctuations to the fluctuations in the system-bath coupling. 
The relationship allows the prediction of decoherence times for 
condensed phase chemical systems from well developed experimental methods.

\vspace{3mm}

\end{abstract} 
 
 
\setlength{\baselineskip}{7mm}     
 
Quantum processes in condensed phases are often studied by focusing  
on a small subset of degrees of freedom and treating the rest as a bath. 
The subsystem of interest may comprise a single molecule, a molecule plus its 
nearest surroundings, or even a single vibrational mode within a molecule. 
The remaining degrees of freedom form the bath. 
In the presence of system-bath interactions, the subsystem's wave function 
evolves into a superposition of quantum states.  
Due to an enormous density of states 
in a macroscopic environment, small differences in the system-bath 
coupling lead to rapid divergence between bath evolutions 
corresponding to different states of the subsystem. 
The reduced density matrix of the microscopic subsystem, 
obtained from the total density matrix by integrating over  
bath degrees of freedom, soon becomes 
diagonal~\cite{Zurek93}.
Quantum states decohere.  
The decoherence rate is determined by the sensitivity of  
bath evolutions to the quantum state of the subsystem.  
Notably, this sensitivity 
also determines bath response 
to a perturbation within the subsystem. 
In the context of the condensed phase chemical physics, the rate of the 
solvent bath rearrangements 
following a perturbation of the solute subsystem  
is described by a well developed solvent response 
theory~\cite{Maroncelli93,Stratt96a,Stratt97}.
In this Letter, we establish for the first time 
a quantitative relationship 
between quantum coherence loss and the short time solvent response. 
 
The new relationship is important because a 
link between solvation dynamics and decoherence is capable of providing 
valuable insights into both phenomena. 
Modern techniques of the solvent response theory employing 
the concepts of dielectric and mechanical relaxation~\cite{Fleming89,FlemingHanggi}, 
and instantaneous normal modes~\cite{Stratt97} then
become transferable to the description of decoherence. 
The theory of short-time solvation can benefit from the 
recent theoretical ideas on quantum Brownian  
motion~\cite{Zurek89,Hu93} 
and quantum measurement~\cite{Diosi95a,Zurek93,Gell-Mann93}, 
where the notion of decoherence appeared first. 
Most importantly, currently available 
experimental means to measure quantum coherence  
loss~\cite{Martin93,Hochstrasser93b,Fleming94} 
and short time solvent response~\cite{Fleming89,Barbara90,Maroncelli93}  
in a non-equilibrium system can be combined. 
A relationship between decoherence and solvation time scales further provides 
a tool to deconvolute the contribution of each effect on observed dynamics. 
The results presented below form 
a basis for evaluation of quantum decoherence times  
in various solute-solvent systems based on the extensive solvent response data 
accessible from both experimental measurements 
and adiabatic molecular dynamics simulations. 
 
Following Ref.~\citen{Simonius78}, 
we consider the direct product of the system $S$  
and bath $B$ Hilbert spaces.  For simplicity, the system space is 
assumed to be two dimensional.  We consider two orthogonal 
states $\phi^S_1$ and $\phi^S_2$ of the system and a system-bath interaction 
that induces quantum transitions in the combined system, with the
bath state responding to that of the system: 
\begin{eqnarray} 
\phi^S_{\alpha}\otimes\phi^B_0 \rightarrow  
\phi^S_{\alpha}\otimes\phi^B_{\alpha},~~\alpha=1, 2, 
\label{eq:transitions} 
\end{eqnarray} 
where $\phi^B_0$ is the initial state of the bath. 
Transitions from an arbitrary initial system state $c_1\phi^S_1 + c_2\phi^S_2$ 
are then described in terms of the reduced density matrix 
\begin{eqnarray} 
\left( \begin{array}{cc} 
       |c_1|^2 & c_1 c_2^* \\ 
       c_1^* c_2 & |c_2|^2\\ 
       \end{array} 
\right) \rightarrow  
\left( \begin{array}{cc} 
       |c_1|^2 & c_1 c_2^* ( \phi^B_2  | \phi^B_1 ) \\ 
       c_1^* c_2 ( \phi^B_1  | \phi^B_2 ) & |c_2|^2\\ 
       \end{array} 
\right). 
\label{eq:dm} 
\end{eqnarray} 
Decoherence is defined~\cite{Zurek93} as decay of the 
non-diagonal matrix elements, which, 
for the reduced density matrix of Eq.~(\ref{eq:dm}), 
is clearly determined by the decay of the inner product of the bath states  
$( \phi^B_1  | \phi^B_2 )$.   
Initially, the bath wave functions coincide: $\phi^B_1 = \phi^B_2 = \phi^B_0$. 
Later on, the bath wave functions correlated with the different states of the system  
diverge, and the overlap integral decreases. 
It is not the decay of the non-diagonal matrix 
elements {\em per se} that is most important from 
the practical point of view, but rather it is
the associated slowing down of quantum  
transitions~\cite{Simonius78,Moore82,Berry95},  
known as the quantum Zeno effect 
in the limit of infinitely fast decoherence~\cite{Berry95}.   
The life time of the quantum state in the presence of a bath 
varies inversely with the decoherence time.  This result follows 
from the Fermi golden rule in the spin-boson model, where the bath is treated 
as a set of harmonic oscillators~\cite{Leggett87,Silbey95},  or 
in the frozen Gaussian formulation~\cite{Nitzan93,Ben96,OPrezhdo97a},
where the bath wave function 
is approximated by a set of Gaussian wave packets~\cite{Heller81}.
Within this Gaussian wavepacket approximation, 
the average decay of the overlap integral and corresponding
non-diagonal matrix elements in Eq.~(\ref{eq:dm}) is described 
by the decoherence function given by Eq.~(39) of Ref.~\citen{OPrezhdo97a} 
\begin{eqnarray} 
D(t) &=& (\phi^B_1|\phi^B_2) = \exp\bigg[-\bigg<\sum_n  
\frac{\Delta F_n^2}{4a_n\hbar^2}\bigg> t^2 \bigg], 
\label{eq:J} 
\end{eqnarray} 
where $\Delta F_n = F_{1n} - F_{2n}$ is the expectation value of the difference 
in the quantum forces experienced by the $n$th bath degree of freedom, and the 
angular brackets indicate thermal averaging. 
We note that here we do not consider the dynamical effect of the bath
on the subsystem energy eigenvalues.  In the presence of bath
induced fluctuations in these values, a standard pure dephasing
contribution (see Refs.~\citen{Borgis95,OPrezhdo97a}) to the decay
of the off-diagonal elements of the reduced density matrix can
potentially contribute as well.
In the adiabatic 
representation the forces are given by the Hellmann-Feynman theorem 
\begin{eqnarray} 
F_{\alpha n} &=& (\phi^S_{\alpha} | \nabla_n H | \phi^S_{\alpha}),~~\alpha=1,2. 
\label{eq:F} 
\end{eqnarray} 
In the low temperature regime
the width $a_n^{-1/2}$ of the Gaussian wave packet equals 
the width of the coherent state of a corresponding
harmonic oscillator, i.e., $a_n=m_n\omega_n/\hbar$. 
For higher temperatures, the width of the wave packet 
incorporates quantum thermal ensemble averaging.    
The thermal width is analytic for harmonic baths 
\begin{eqnarray} 
a_n &=& \frac{m_n\omega_n}{\hbar} \tanh\left(\frac{\hbar\omega_n}{2k_b T}\right). 
\label{eq:tanh} 
\end{eqnarray} 
For arbitrary baths, the width can be defined via the thermal 
de Broglie wave length $\lambda_{B} = (2\pi\hbar^2/mk_BT)^{1/2}$. 
An alternative expression for the
thermal width is derived in Ref.~\citen{Nitzan93}, Sec.~IV 
by comparing the exact and Gaussian wave packet results for the transition 
rate in a double well system 
\begin{eqnarray} 
\label{eq:A} 
a_n &=& \frac{m_n\omega_n}{\hbar} A_n, \\ 
A_n &=& \left[\coth\left(\frac{\hbar\omega_n}{2k_BT}\right) 
-\frac{2k_BT}{\hbar\omega_n}\right]^{-1}. \nonumber 
\end{eqnarray} 
This expression reduces to the coherent state width in the low temperature 
case, and gives 
\begin{eqnarray} 
a_n &=& \frac{6m_nk_BT}{\hbar^2} \simeq \left(\frac{\lambda_{B}}{6}\right)^{-2}, 
\label{eq:width} 
\end{eqnarray} 
in the high temperature limit.  The last formula is particularly useful, since 
it yields a width which is independent of the frequency;  Eq.~(\ref{eq:width}) 
is designed~\cite{Nitzan93} for use in molecular dynamics simulations,  
where thermal averaging 
over bath states is performed classically. 
 
Turning to solvation dynamics, 
the response of the solvent bath to a quantum transition
within the solute subsystem 
is quantified by the normalized correlation function $C$ of the energy gap $U$ 
between the two quantum states~\cite{Fleming89,Stratt94a}. 
The fluctuation-dissipation theorem relates the non-equilibrium solvent response 
to the regression of fluctuations $\delta U$ of the gap $U$ in equilibrium 
\begin{eqnarray} 
C(t) &=& \frac{\langle \delta U(t) \cdot \delta U(0)\rangle} 
{\langle (\delta U)^2 \rangle}. 
\label{eq:C} 
\end{eqnarray} 
The short time solvation dynamics that is of relevance 
in the present discussion
depends solely on the change in the solute-solvent  
coupling due to the quantum transition. 
The microscopic {\it short time} expression for $C(t)$ 
has been obtained in Ref.~\citen{Stratt94a} [Eqs.~(2.18),~(2.19)]  
by expanding $C(t)$ in a set of independent modes and in time, yielding
\begin{eqnarray} 
C(t) &=& \exp\left[-\frac{k_BT} {2\big<(\delta U)^2\big>}  
\left< \sum_n (U'_n)^2 \right> t^2\right], 
\label{eq:C1} 
\end{eqnarray} 
where $U'_n$ is the derivative of $U$ with respect to the $n$th mass-weighted  
solvent coordinate 
\begin{eqnarray} 
U'_n &=& m_n^{-1/2} \frac{d U}{d x_n} = m_n^{-1/2} \Delta F_n. 
\label{eq:U'} 
\end{eqnarray} 
 
The decoherence $\tau_{_D}$ and Gaussian solvation $\tau_g$ time scales 
are given by the variances of the decoherence $D(t)$ and solvent response $C(t)$  
functions of Eqs.~(\ref{eq:J}) and~(\ref{eq:C1}), respectively.  
The structure of  
the equations is clearly similar, and by comparison we obtain 
\begin{eqnarray} 
\left[\frac{\tau_{_D}}{\tau_g}\right]^2 &=&  
\frac{2k_BT}{\big<(\delta U)^2\big>} 
\frac{\sum_n \Delta F_n^2(m_n)^{-1}}{\sum_n \Delta F_n^2 (a_n\hbar^2)^{-1}}.
\label{eq:main0} 
\end{eqnarray} 
With the high temperature limit expression for the width [Eq.~(\ref{eq:width})] 
the formula simplifies to
\begin{eqnarray} 
\left[\frac{\tau_{_D}}{\tau_g}\right]^2 &=&  
\frac{12 (k_BT)^2}{\big<(\delta U)^2\big>} = 
\frac{6 k_BT}{\lambda}. 
\label{eq:main} 
\end{eqnarray} 
Here, $2\lambda$ is the Stokes' shift, 
the difference between the equilibrium absorption and 
emission maxima. Within a linear response regime, $\lambda$ 
is related~\cite{Silbey95} 
to the fluctuations in the quantum energy 
gap by the fluctuation-dissipation theorem: 
$\big<(\delta U)^2\big> = 2\lambda k_BT$. 
Eq.~(\ref{eq:main}) establishes direct proportionality 
between the short time evolution of decoherence and of solvent response
in the high temperature limit for the bath.
The proportionality coefficient is determined by the ratio of the thermal 
energy fluctuations $(k_BT)^2$ to the fluctuations in the system-bath 
coupling $\big<(\delta U)^2\big>$.  The letter can be expressed in terms
of the Stokes' shift in the linear response regime. We note that
the dependences are sensible:
High temperatures accelerate solvation dynamics,
large equilibrium fluctuations in the coupling, as well as large 
Stokes' shifts, are indicative of fast decoherence. 

We note that the relationship between the decoherence 
and solvation time scales presented 
above using the Gaussian wave packet approximation for the bath wave function
also necessarily pertains to the spin-boson model. The spin-boson 
Hamiltonian describes a two-level system linearly coupled to a harmonic 
bath~\cite{Leggett87} 
\begin{eqnarray} 
H_{SB} &=& -\frac{1}{2}\hbar\Delta\sigma_x + \frac{1}{2}U_0 \sigma_z 
+ \sum_n\left(\frac{1}{2} m_n \omega_n x^2_n 
+ \frac{p^2_n}{2m_n} \right) + \frac{1}{2} q_0\sigma_z \sum_n 
c_n x_n, 
\label{eq:HSB} 
\end{eqnarray} 
where, $U_0 $ and $q_0$ are the energy and coordinate displacements 
between the pair of potential minima, $\Delta$ is the intrinsic 
coupling between the two quantum states, $c_n$ is the system-bath 
coupling constant, and $\sigma_z$, $\sigma_x$ are the Pauli matrices.   
The terms containing $\sigma_z$ describe the energy gap
\begin{eqnarray} 
U_{SB} &=& U_0 + q_0 \sum_n c_n x_n. 
\label{eq:USB} 
\end{eqnarray} 
According to Eqs.~(\ref{eq:C1}),~(\ref{eq:U'}),  
the short-time solvent response function of the spin-boson model is  
\begin{eqnarray} 
C_{SB} &=& \exp\bigg[-\frac{1}{2}\frac{k_BT}{\big<(\delta U)^2\big>} 
\sum_n \frac{q_0^2 c_n^2}{m_n} t^2\bigg]. 
\label{eq:SSB} 
\end{eqnarray} 
The decoherence function can be extracted from the 
Fermi golden rule result for the spin-boson problem [Eqs.~(3.35),~(3.36)  
and~(3.2) of Ref.~\citen{Leggett87}] 
\begin{eqnarray} 
D_{SB} &=& \exp\bigg[-(q_0^2/\pi\hbar)Q_2(t)\bigg], 
\label{eq:DSB} 
\end{eqnarray} 
with 
\begin{eqnarray} 
Q_2(t) &=& \int_0^{\infty} \frac{J(\omega) ( 1 - \cos\omega t )}{\omega^2} 
\coth(\hbar\omega/2k_BT) d\omega, 
\label{eq:Q2} 
\end{eqnarray} 
and  
\begin{eqnarray} 
J(\omega) &=& \frac{\pi}{2} \sum_n \frac{c_n^2}{m_n\omega_n} \delta(\omega-\omega_n). 
\label{eq:Jw} 
\end{eqnarray} 
The short-time expansion of $Q_2(t)$ gives 
\begin{eqnarray} 
D_{SB} &=& \exp\bigg[-\frac{1}{4}\sum_n \frac{q_0^2 c_n^2}{m_n\hbar\omega_n} 
\coth(\hbar\omega_n/2k_BT) t^2\bigg] \\ 
&=& \exp\bigg[-\frac{1}{4}\sum_n \frac{q_0^2 c_n^2}{a_n\hbar^2} t^2 \bigg], 
\label{eq:D1SB} 
\end{eqnarray} 
with $a_n$ as in Eq.~(\ref{eq:tanh}).   
The expression in Eq.~(\ref{eq:D1SB}) is a specific case of 
Eq.~(\ref{eq:J}) with $\Delta F_n=-q_0c_n$. 
Comparison of Eqs.~(\ref{eq:SSB}) and~(\ref{eq:D1SB}) 
leads to the relationship given in Eq.~(\ref{eq:main0}) between 
the decoherence and Gaussian solvation times. 
 
We can directly test Eq.~(\ref{eq:main})  for the case of relaxation
following transition
from the first excited to the ground state of the hydrated electron, whose 
time dependent properties are well studied 
theoretically (see Refs.~\citen{Nitzan93,Ben94,OPrezhdo97a,Ben96,OPrezhdo96} 
and references therein). 
Based on a molecular dynamics trajectory for the first excited 
state of the hydrated electron~\cite{Ben94}, the equilibrium energy
gap $U$ is 0.56~eV and 
the fluctuation in the energy gap
$\big<(\delta U)^2\big>^{1/2}$ is 0.21~eV, 
which corresponds to a 1.7~eV Stokes' shift (2$\lambda$) at room
temperature.  This Stokes' shift corresponds closely to that found
in non-adiabatic simulations~\cite{Ben94d}.
The short-time  component of the solvent response function $C(t)$
of Eq.~(\ref{eq:C}) is found to be
characterized by a 10.6~fs Gaussian time scale, $\tau_g$.
From these data, 
the decoherence time $\tau_{_D}$ of the first excited state of 
the hydrated electron 
calculated via Eq.~(\ref{eq:main}) is 4.5~fs.
This estimate 
falls within the previously reported
range of 2.7-5.1~fs~\cite{Nitzan93,OPrezhdo97a,Ben96},
with the value of 5.1~fs obtained in the high temperature approximation
for Eq.~(\ref{eq:A}).
 
The properties of the solute and the nature of the quantum transition  
define the difference in the solute-solvent potential for the initial 
and final states and, therefore, determine the magnitude of the Stokes' shift 
in the relationship~(\ref{eq:main}) between the decoherence 
and solvation times. 
Intrinsic solvent properties also, to some extent, 
affect the magnitude of the Stokes' shift. 
However, the major solvent influence on the duration of quantum coherence  
is due to solvent's ability to respond to a perturbation in the solute,
i.e. as reflected in the rate of solvent response, $\tau_g$.
Based on the success above for the hydrated electron, one can
address other systems, which will demonstrate the variability of
decoherence times.  For a styryl dye in methanol~\cite{Ernsting95},
experiment yields an estimated $\tau_g$ of 40~fs and a Stokes' shift
of 115~nm yielding a decoherence time of 6.8~fs.
Acetonitrile [CH$_3$--CN], the next solvent in a logical
series exhibits a 100~fs  
experimental short time 
solvent response~\cite{Fleming91}. 
Simulation of the electronic transition of the betaine-30 molecule 
in acetonitrile is characterized by a similar 91~fs short-time 
solvation~\cite{Lobaugh97}.  
The value for $\tau_g$, together with the 0.16~eV 
electronic energy gap fluctuation evaluated along the ground state 
trajectory~\cite{Lobaugh97},
leads via Eq.~(\ref{eq:main}) to a substantially longer 49~fs decoherence time. 
Compared to the protic solvents, acetonitrile is much less effective 
in destroying quantum coherence. 

In summary, we have presented an analytical relationship between
the time scale for quantum decoherence and that governing the short
time response of solvent to a perturbation in the solvent-solute
coupling.  The proportionality constant relating these requires
knowledge of only the Stokes' shift associated with the change in
solute state.  The expression successfully reproduces results obtained
directly by other routes, allows the prediction of decoherence times
for other solution systems.  It is expected that since the required
input data is becoming readily accessible experimentally for even
the most rapidly responding condensed phase environments, the derived
relation will be very valuable to advancing the study of both
decoherence and condensed phase chemical dynamics more generally.

\vskip5mm

{\bf Acknowledgment:} The authors are grateful to the National Science
Foundation (CHE-9314066) for support of the research reported here.
OVP acknowledges the support of the Hemphill/Gilmore fellowship fund.

 
\newpage 
 
 
 

\end{document}